\documentclass[a4paper]{jpconf}
\usepackage{graphicx}
\usepackage{amsmath}
\usepackage{multirow}
\usepackage{upgreek}
\usepackage{subcaption}
\usepackage{siunitx}					
\usepackage[colorlinks=true, a4paper=true, pdfstartview=FitV,
linkcolor=black, citecolor=black, urlcolor=black]{hyperref}
\usepackage{url}
\usepackage{scalerel}

\DeclareFontFamily{U}{BOONDOX-calo}{\skewchar\font=45 }
\DeclareFontShape{U}{BOONDOX-calo}{m}{n}{
  <-> s*[1.05] BOONDOX-rcalo}{}
\DeclareFontShape{U}{BOONDOX-calo}{b}{n}{
  <-> s*[1.05] BOONDOX-b-calo}{}
\DeclareMathAlphabet{\mathcalboondox}{U}{BOONDOX-calo}{m}{n}
\SetMathAlphabet{\mathcalboondox}{bold}{U}{BOONDOX-calo}{b}{n}
\DeclareMathAlphabet{\mathbcalboondox}{U}{BOONDOX-calo}{b}{n}

\begin{document}

\title{The E302 instability-versus-efficiency experiment at FACET-II}

\author{O G Finnerud \textsuperscript{1}, C A Lindstrøm\textsuperscript{1}, J B B Chen\textsuperscript{1} and E Adli\textsuperscript{1}}

\address{$^1$ Department of Physics, University of Oslo, Oslo, Norway}

\ead{o.g.finnerud@fys.uio.no}

\begin{abstract}
 We discuss plans for the E302 instability--efficiency experiment, starting in 2024 at the recently upgraded FACET-II facility at SLAC National Accelerator Laboratory. The beam-breakup instability will be the main area of study for the E302 experiment. With the imaging spectrometer at FACET-II, we introduce a novel technique for observing and quantifying the amplitude growth of the trailing bunch due to the transverse instability. Using the transverse position on the spectrometer screen and the transfer matrix of the magnetic lattice used for the spectrometer, we aim to extract a $x'$--$E$ charge distribution that can be used to quantify the amplitude of the beam. By varying the trailing bunch's charge and, hence, the beam loading of the accelerating field, we aim to adjust the wake-to-beam power transfer efficiency in the E302 experiment. We plan to quantify the amplitude for different configurations of the beam charge and, hence, investigate the relationship between the beam-breakup instability and efficiency. We use a combination of particle-in-cell (PIC) codes to simulate a beam-driven plasma wakefield accelerator from start-to-end with a FACET-II-like spectrometer and demonstrate the methodology that will be used for the instability studies at the E302 experiment.

\end{abstract}
\section{Introduction}
Plasma wakefield acceleration (PWFA) \cite{tajima, chen} promise to reduce the length and hence construction costs of new particle accelerators significantly as they are capable of producing accelerating gradients of the order 10--100 GV/m by far exceeding the 100 MV/m limit in radio frequency (RF) accelerators. This is particularly promising for the development of new $\mathrm{e^{+}}$--$\mathrm{e^{-}}$ colliders aimed at reaching hundreds of GeV \cite{halfh}, but also relevant for other applications such as free-electron lasers (FEL) \cite{FEL}.

In recent years, the PWFA community has reached several milestones for the development of a plasma linac such as multi-GV/m gradients with high efficiency and emittance preserving high-efficiency acceleration with low final energy spread for double-bunch acceleration \cite{litos, energyspreadpreservedhigheffic}. Acceleration of a large amount of charge with high efficiency over a substantial distance remains to be demonstrated. Small emittance is particularly important for colliders to ensure high luminosity per wallplug-power. Therefore, emittance preservation is crucial for good beam quality in applications for FELs and colliders. The beam-breakup instability \cite{Lebedev} is an effect where an offset beam alters the trajectory of the sheath electrons, which induces short-range transverse wakefields. The presence of these wakefields in the ion-focusing channel constitutes a driven harmonic oscillator that resonantly drives an amplitude growth of the beam from head to tail. This instability is an obstacle to simultaneously achieving high efficiency and emittance preservation in PWFAs. Recently, it has been suggested that the beam-breakup instability imposes a fundamental limit on the acceleration efficiency achievable in plasma accelerators. This instability, seeded by a misalignment of the trailing and driving bunches, has a detrimental effect on the transverse phase-space density of the trailing bunch if not mitigated. Recent work using particle-in-cell (PIC) codes has shown that the relation between efficiency and amplitude growth due to the beam-breakup instability may not be as detrimental as initially expected \cite{BensPhD}. However, testing the predictions experimentally is needed to quantify the size of the effect and determine appropriate mitigation strategies for future PWFA accelerators.

This is the main goal of the E302 experiment at SLAC, which will be conducted at the recently upgraded Facility for Advanced Accelerator Experimental Tests (FACET-II) \cite{Joshi_2018facet}. The E302 experiment will measure the efficiency of the accelerated bunch, quantify the effect of transverse instabilities and compare this to theoretical predictions. Subsequently, we will also experimentally test mitigation methods.

\section{The E302 experiment}
The first goal for E302 is to implement a method that can quantify the amplitude growth of the trailing bunch due to the beam-breakup instability. For the long-term plan, we will also aim to investigate the effect of efficiency on amplitude growth due to the transverse instability and determine the effectiveness of instability mitigation strategies. This section describes the FACET-II facility, relevant diagnostics and the methodology we will employ during the experiment. This experiment builds on the methods developed and the results from previous experiments at FACET \cite{ADLItransverseosc}.

\subsection{FACET-II facility}

\begin{figure}[h!]
    \includegraphics[scale=0.23]{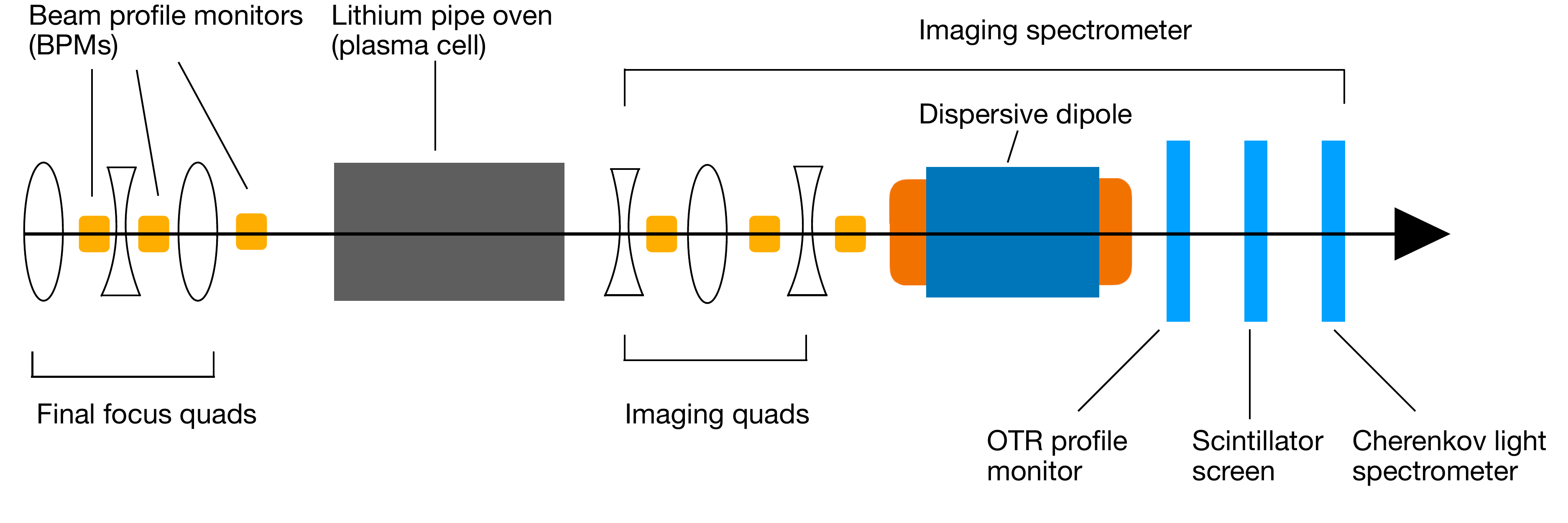}
    \caption{Sketch of the FACET-II beamline from the final focus to the electron diagnostics relevant for the E302 experiment. The black line represents the beam centroid trajectory.}
    \label{beamline}
\end{figure}

The FACET-II facility utilises the middle kilometre of the SLAC linac and is situated between LCLS-II and LCLS. The new RF-photocathode injector employed in FACET-II is capable of producing electron beams with a normalised emittance of a few \textmu$\mathrm{m}$ \cite{storey2023wakefield}. A simplified sketch of the beamline from the final focus lattice to the electron diagnostics is shown in Fig.~\ref{beamline}.

If calibrated well and imaged correctly, the spectrometer will extract each beam's charge, energy spectra and overall efficiency. The spectrometer will be our main diagnostic used for the analysis. Using the $x$--$E$ charge distribution on the spectrometer screen, we will extract an oscillation amplitude from the beam-breakup instability. By measuring the charge and energy of the bunches before and after acceleration, we will quantify the wake-to-trailing bunch efficiency. Understanding the imaging of the spectrometer and doing Gaussian fits on the images produced at the E302 experiment will be important to separate beam-plasma effects from imaging effects. The final focusing system consists of a quadrupole triplet. A plasma column is created from lithium vapour in a pipe oven \cite{pipeoven} with Helium gas acting as a buffer. Following the oven, another quadrupole triplet focuses the beam on the spectrometer screen.
A dipole magnet then disperses the beam vertically such that its energy spectrum can be resolved at the electron beam diagnostics. The three relevant diagnostic screens for the E302 experiment that we can use to observe the spectrally dispersed transverse effects are the in-vacuum OTR, scintillation screen and Cherenkov spectrometer \cite{ADLIcherenkov}. The Cherenkov spectrometer can be used for the broadband spectrum of the beams, the scintillator screen can be used for high-resolution imaging of the transverse structure of low charge density beams and the OTR screen can be used for high-resolution imaging of the transverse structure of high charge density beams. Additionally, toroidal current transformers will be used for high-precision measurement of total charge. 

\subsection{Methodology}
We introduce a single-shot concept for quantifying the amplitude growth of the trailing bunch due to transverse instabilities in the E302 experiment. The transverse position on the screen $x_{\mathrm{screen}}$ for each beam slice will have contributions from the initial transverse offset in the $x$--plane $x_{0}$ and initial angle $x'_{0}$ at the point where it was imaged, e.g., at the plasma exit as 
\begin{equation}
    x_{\mathrm{screen}} = m_{11}(E)x_{0} + m_{12}(E)x_{0}',
    \label{Rmatrixbasic}
\end{equation}
where $m_{11}(E)$ and $m_{12}(E)$ are the transfer matrix elements of the imaging-spectrometer lattice. The effective field strength of the quadrupoles depends on the energy of the incoming beam slices; therefore, these slices will appear spread out in the $x$--plane of the spectrometer due to either under-focusing or over-focusing, forming a ``butterfly" pattern on the screen. The further away we are from the imaging energy (i.e., $m_{12}(E) = 0$) in the $y$--plane, the more the $m_{12}(E)x_{0}'$ term will dominate Eq~\ref{Rmatrixbasic}. We propose using this term and the chromaticity of the quadrupoles to measure a $x'$--$E$ charge distribution. The points on the spectrometer with the largest transverse amplitude represent the oscillation amplitude where the unmeasured $x_{0}$ has the smallest contribution. The value of $m_{12}(E)$ for the imaging energy is set by the magnetic lattice design. Because the calibrated spectrometer gives us the energy spectrum of the beam, we can calculate the values of $m_{12}(E)$ for beam slices with different energies using the effective field strength experienced by a beam slice. Because of the strong focusing forces in a plasma accelerator, the beam size is small and divergence is large, the angular contribution will dominate over the positional contribution in Eq~\ref{Rmatrixbasic} at all energies except close to the imaging energy (where $m_{12}(E)$ = 0). This means we can approximate the initial angle of the particles using:
\begin{equation}
x'_{0} = \frac{x_{\mathrm{screen}}}{m_{12}(E)}.    
\label{anglemethod}
\end{equation}
We can then use $x'_{0}$ to measure the oscillation amplitude. Typically, when working with amplitudes of oscillations in plasma-driven accelerators, one employs normalised coordinates to account for the focusing and acceleration of the beam. The position term will act as an uncertainty term: \begin{equation}
\sigma_{x', \mathrm{error}} = \frac{\sigma_x m_{11}(E)}{m_{12}(E)},
\end{equation}
where $\sigma_x$ is the approximate beam size.
We use the transformation matrix derived in \cite{beta_mismatch} to transform $x'_{0}$ to the normalised angle $x'_{0n}$ as
\begin{equation}
    x'_{0n} = \sqrt{\gamma\beta} x'_{0}.
    \label{normaliseddiv}
\end{equation}

If there is no instability observed, consistent with previous studies at FACET \cite{ADLItransverseosc}, we expect the normalised amplitude of the trailing bunch to stay constant. If, on the other hand there is an instability, we expect to see a growth in normalised amplitude from the head to the tail. By comparing configurations of high and low-charge trailing bunches, we can investigate the effect of beam loading, hence efficiency, on the effect of transverse instabilities.

\subsection{Simulating the experiment}
We simulate a beam-driven plasma wakefield accelerator from start to end, using high and low-charge trailing bunches with FACET-II-like parameters on a FACET-II-like spectrometer. Ion motion is currently excluded in these simulations. However, it will be included in upcoming simulations and subsequent analyses. The parameters for the simulations are given in Table \ref{beamparameters}, where both simulations have the same parameters, only differing in total charge of the trailing bunch. We have increased the driver energy to avoid head erosion and added an initial offset to the trailing bunch of 1 \textmu$\mathrm{m}$ in the $x$--plane. We employ a long trailing bunch in both cases such that we see a detailed evolution of the beam oscillation amplitude from the head of the trailing bunch to its tail. 
 
\begin{figure}[h!]
  \centering
  \begin{minipage}[b]{0.46\textwidth}
    \includegraphics[width=\textwidth]{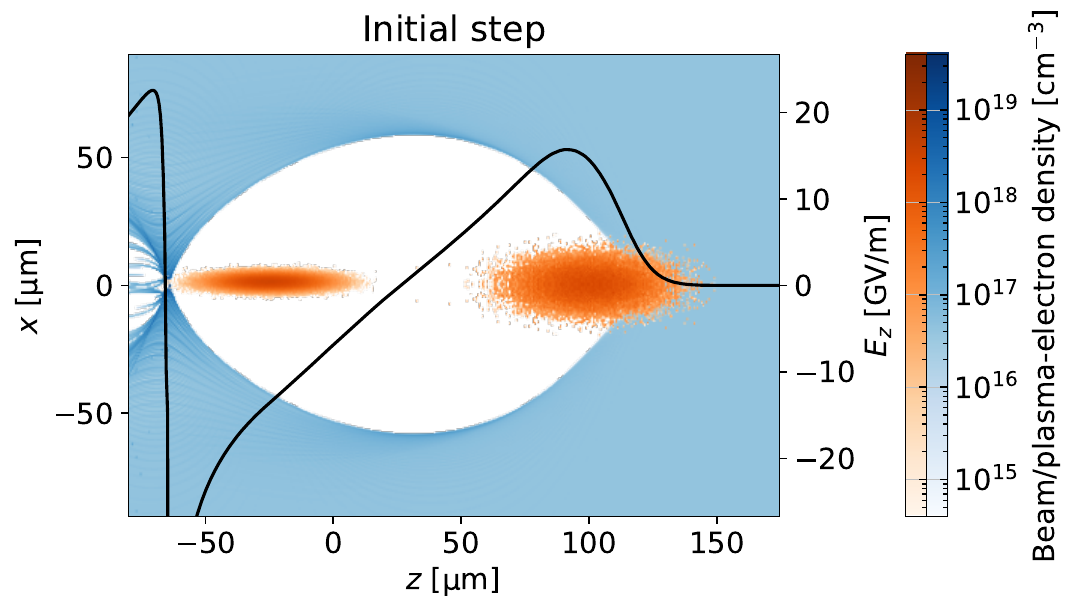}
    \caption{Initial electron density for the bunches and plasma with a fully blown out bubble with an underloaded accelerating field $E_{z}$.}
    \label{understartt}
  \end{minipage}
  \hfill
  \begin{minipage}[b]{0.46\textwidth}
    \includegraphics[width=\textwidth]{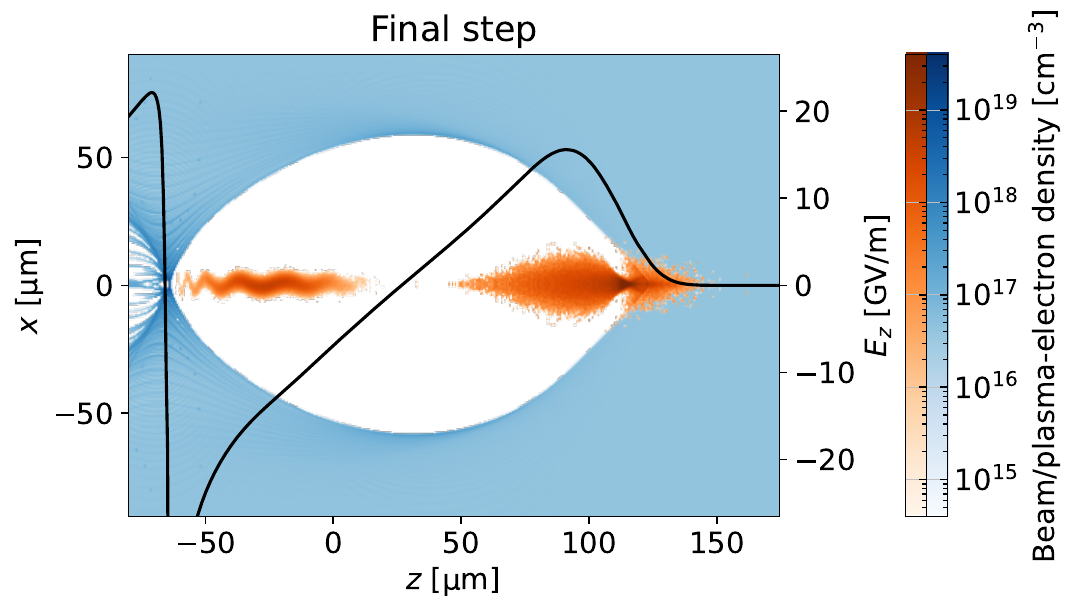}
    \caption{Final electron density for the bunches and plasma with a fully blown out bubble with an underloaded accelerating field $E_{z}$.}
    \label{underfinal}
  \end{minipage}
\end{figure}

 \begin{table}[h!]
\centering
  \renewcommand{\arraystretch}{1.5} 
\begin{tabular}{l c c}
  \multirow{2}{*}{} & \multicolumn{2}{c}{} \\
  
  & Drive beam & Main beam \\
  \hline
  $E$ [GeV] & 100* & 10 \\
  \hline
  $Q$ (nC) & -1.6 & -1.5/-0.2 \\
  \hline
  $\sigma_z$ (\textmu m) & 13 & 10 \\
  \hline
  $\epsilon_n$ (mm-mrad) & 10 $\times$ 10 & 10 $\times$ 10 \\
  \hline
  $\beta$ (cm) & 50 $\times$ 50 & 0.53 $\times$ 0.53 \\
  \hline
  dE/E (\%) & 0.15 & 0.5\\
  \hline
\end{tabular}
\caption{Simulation parameters used in this analysis. *We increase the energy of the driving bunch to avoid head erosion.}
\label{beamparameters}
\end{table}
We use the multipurpose code OCELOT \cite{OCELOTE} to simulate the spectrometer and the 3D quasi-static code HiPACE++ \cite{HiPACE++} to simulate the plasma interaction. In both the underloaded and overloaded cases, we image a little below the tail to ensure we can see most of the beam and that $m_{12}(E)x_{0}'$ is large enough to dominate the transverse position on the screen. In Figs.~\ref{understartt} and \ref{underfinal}, we show the electron density distributions of the plasma and beams for the underloaded case. We also indicate the value of the on-axis accelerating field $E_{z}$.

\begin{figure}[h!]
    \centering
    \includegraphics[scale = 0.79]{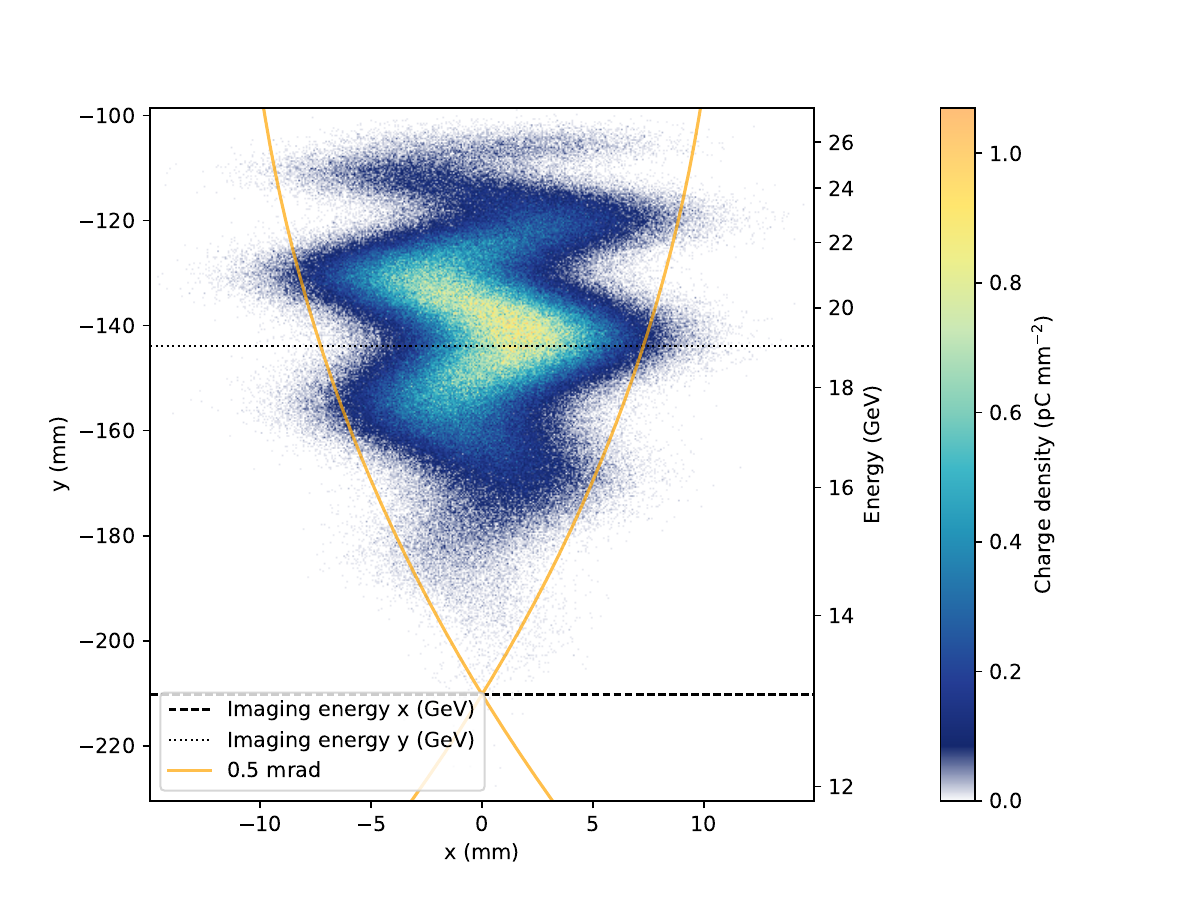}
    \caption{Simulated FACET-II spectrometer screen for an underloaded wakefield and a trailing bunch with an initial offset of 1 \textmu$\mathrm{m}$ in the $x$--plane. The orange lines represent a divergence of 0.5 mrad.}
    \label{underloaded}
\end{figure}

\begin{figure}[h!]
    \centering
    \includegraphics[scale = 0.79]{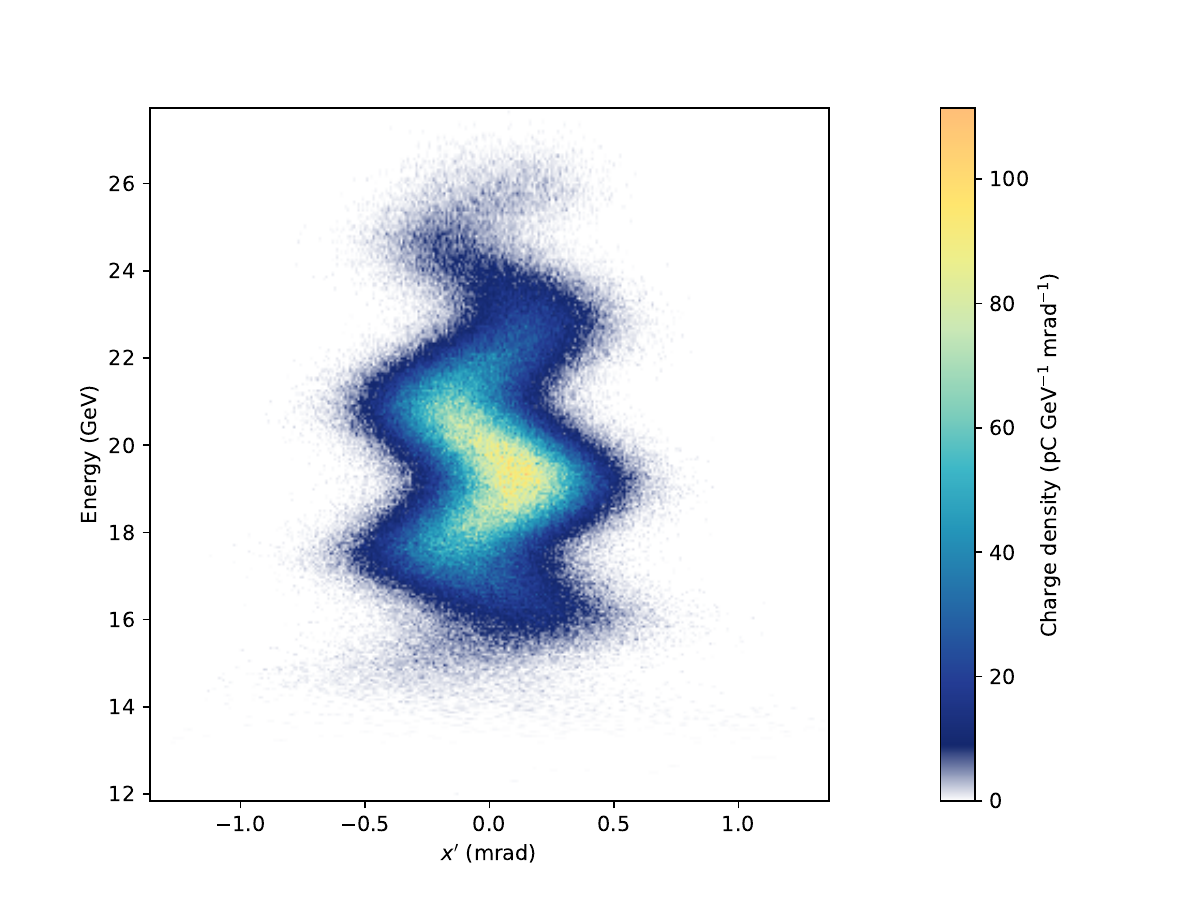}
    \caption{Mapping of the $x'$--$E$ charge distribution from the spectrometer image for an underloaded wakefield and a trailing bunch with an initial offset of 1 \textmu$\mathrm{m}$ in the $x$--plane.}
    \label{underscaled}
\end{figure}

From these figures, we see that the accelerated trailing bunch will have higher energy towards its tail. We can also see some low-amplitude oscillations of the trailing bunch in Fig.~\ref{underfinal}. However, further analysis is needed to determine if this indicates transverse instabilities or energy-dependent betatron oscillation. In Fig.~\ref{underloaded}, we can see the simulated spectrometer screen for the underloaded case.
On the spectrometer image, there are clear oscillations of the beam centroid in $x$ from the head of the beam to the tail. This is caused by the beam oscillating around its initial offset in the focusing channel. Typically, to see an indication of transverse instabilities, one would observe the oscillation amplitude increase towards the tail. To better understand the beam's oscillations in the plasma channel and as a prerequisite to measuring an oscillation amplitude, we transform the spectrometer image to a $x'$--$E$ image. We do this by scaling the pixels to the value of $x_{0}' = x_{\mathrm{screen}}/m_{12}(E)$ and interpolating on a newly defined $x'$ axis. This is shown in Fig.~\ref{underscaled}.
The oscillation amplitude appears roughly constant along the beam. Hence, there is no clear indication of instabilities, and the oscillations shown on the spectrometer image are likely due to the energy spread longitudinally in the trailing bunch that causes longitudinal slices to oscillate at different betatron frequencies. The extracted wake-to-trailing bunch efficiency from the simulation is 14.2\%.

\begin{figure}[!h]
  \centering
  \begin{minipage}[b]{0.46\textwidth}
    \includegraphics[width=\textwidth]{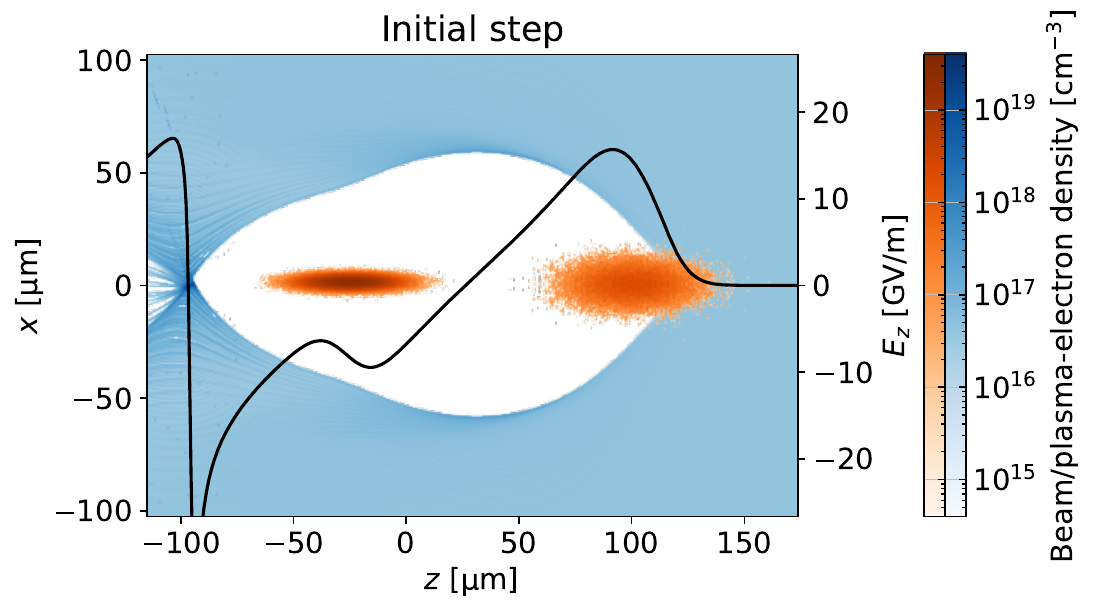}
    \caption{Initial electron density for the beams and plasma with a fully blown out bubble with an overloaded accelerating field $E_{z}$.}
    \label{overstartt}
  \end{minipage}
  \hfill
  \begin{minipage}[b]{0.46\textwidth}
    \includegraphics[width=\textwidth]{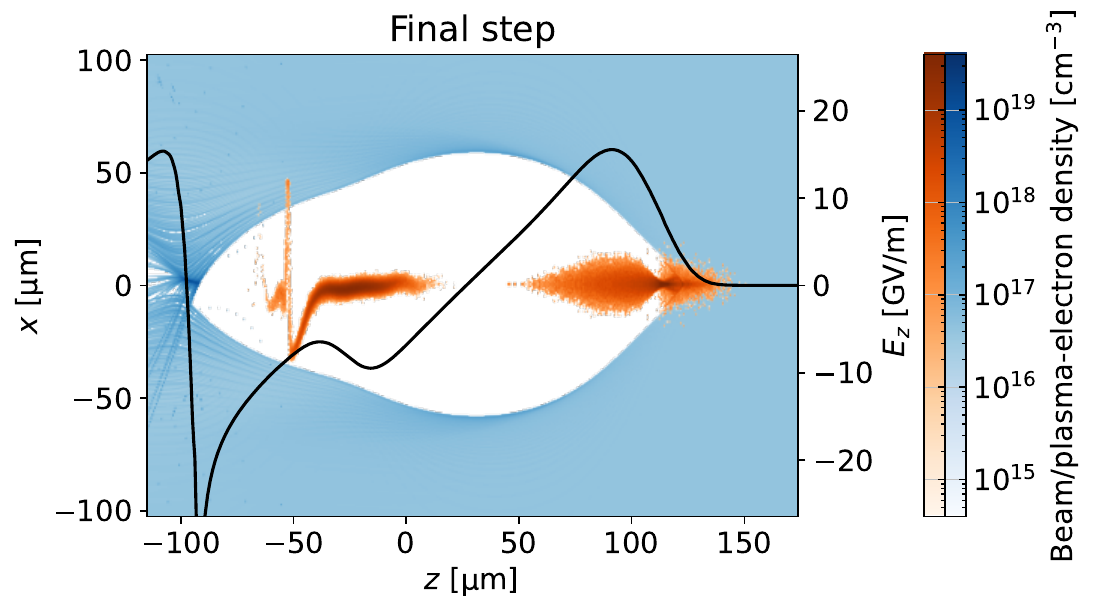}
    \caption{Final electron density for the beams and plasma with a fully blown out bubble with an overloaded accelerating field $E_{z}$.}
    \label{overfinal}
  \end{minipage}
\end{figure}

\begin{figure}[h!]
    \centering
    \includegraphics[scale = 0.79]{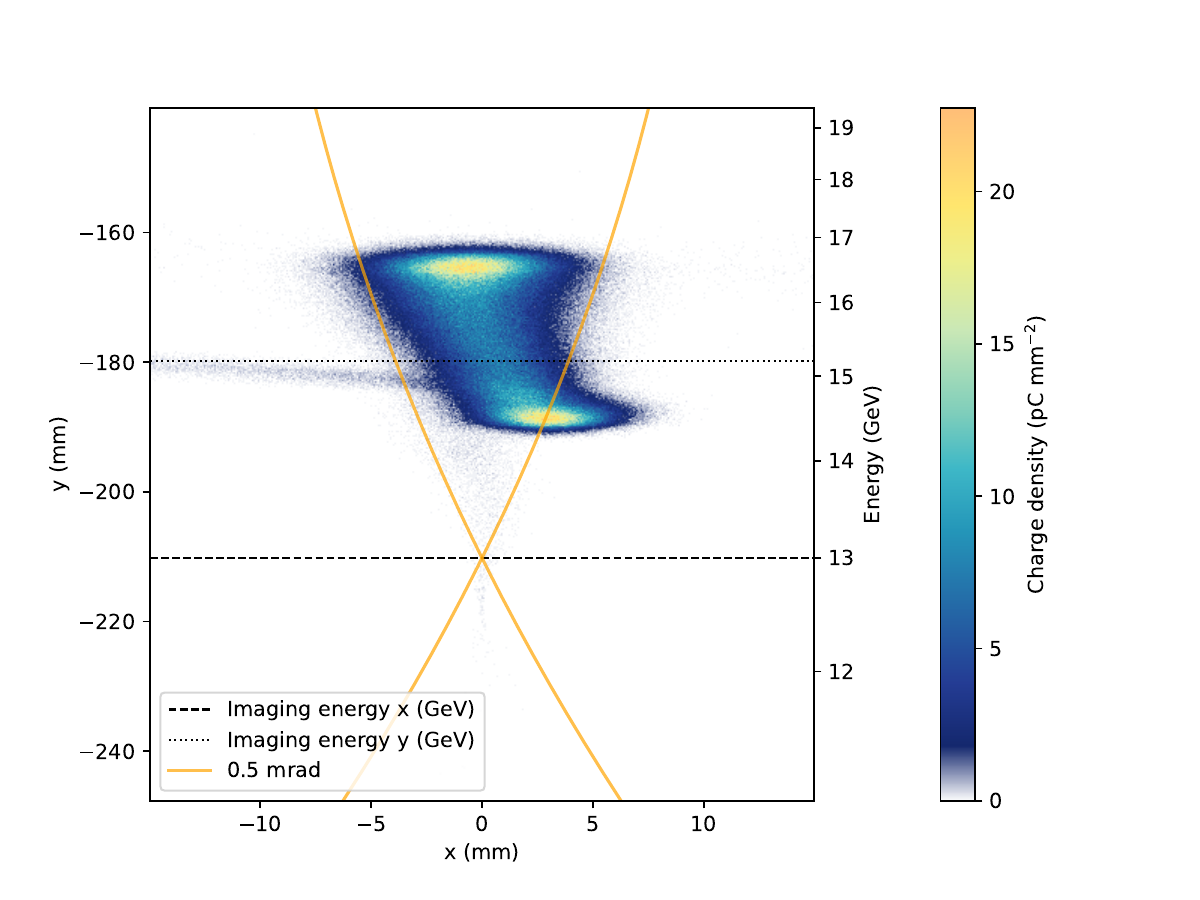}
    \caption{Simulated FACET-II spectrometer screen for an overloaded wakefield and a trailing bunch with an initial offset of 1 \textmu$\mathrm{m}$ in the $x$--plane. The orange lines represent a divergence of 0.5 mrad.}
    \label{overspec}
\end{figure}

\begin{figure}[h!]
    \centering
    \includegraphics[scale = 0.79]{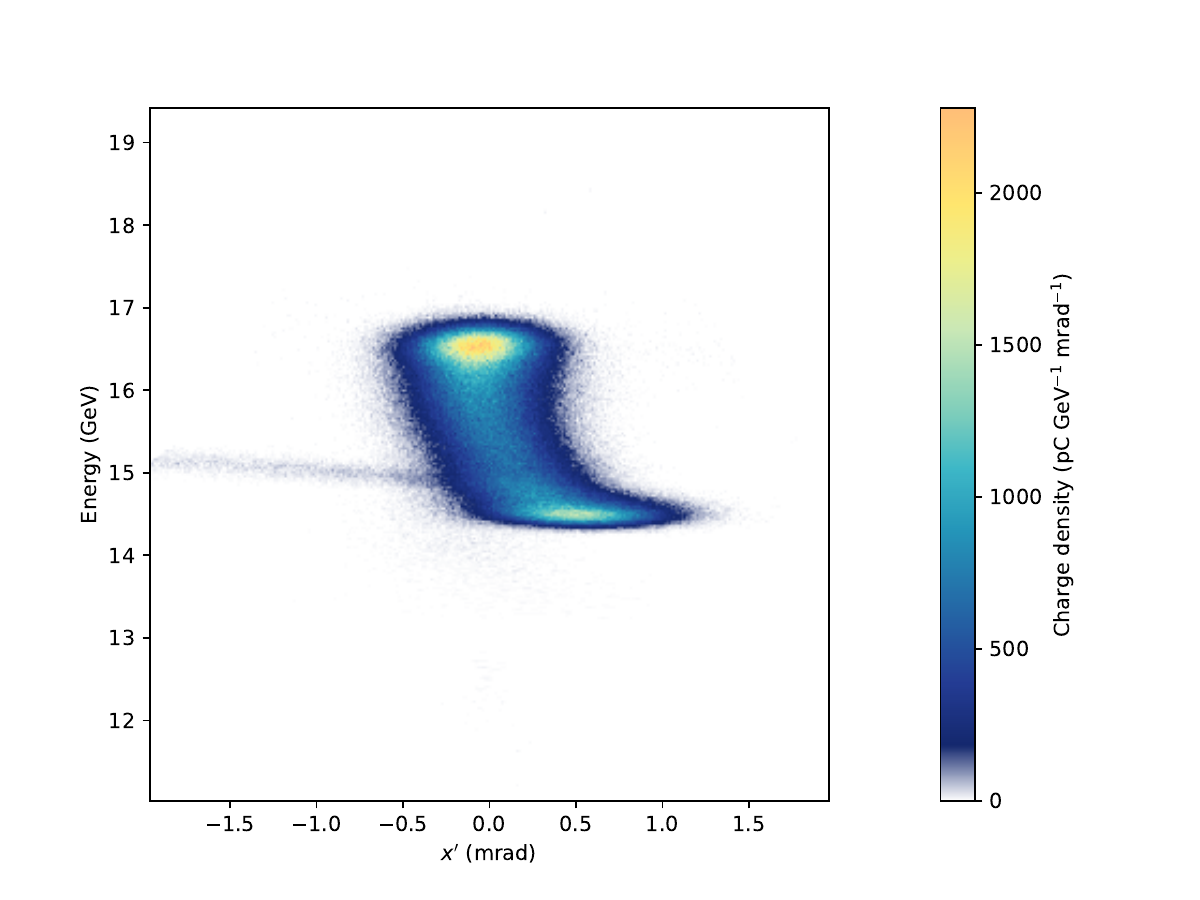}
    \caption{Mapping of the $x'$--$E$ charge distribution from the spectrometer image for an overloaded wakefield and a trailing bunch with an initial offset of 1 \textmu$\mathrm{m}$ in the $x$--plane.}
    \label{overscaled}
\end{figure}

The beam and plasma electron density for the overloaded case at the start and end of the plasma channel can be seen in Figs.~\ref{overstartt} and \ref{overfinal}.
As can be seen from the accelerating field strength indicated, the energy of the accelerated bunch will increase from the tail towards the head of the bunch, as the head will experience a larger $E_{z}$ over the acceleration length. We also note that the beam density distribution at the end of the simulation shows a clear indication of transverse instabilities. We can see that towards the back of the bubble, longitudinal slices are deflected towards and past the plasma electron sheath. If we compare the underloaded and overloaded cases, the beam density distribution indicates the presence of transverse instabilities more clearly for the overloaded case than the underloaded, as expected. The spectrometer screen for the overloaded case can be seen in Fig.~\ref{overspec}.
One can see from the butterfly drawn that a part of the tail appears deflected away from the expected position enclosed by the butterfly. This might indicate the presence of instabilities. The scaled $x'$--$E$ charge distribution is shown in Fig.~\ref{overscaled}.
From the $x'$--$E$ charge distribution, we can see a beam roughly constant in $x'$ at high energy. However, towards lower energies, which in this case represents the tail of the beam, we see that slices are deflected towards large $x'$. This is consistent with the density distribution from Fig.~\ref{overfinal} and can indicate the presence of transverse instabilities.  We also note that these slices extend past the range covered in the underloaded case and past one mrad. The extracted wake-to-trailing bunch efficiency from the simulation is 61.9\%.

\subsection{Mitigation techniques}
If we reduce the beam density at E302, we can operate in the quasi-linear regime, where there are still electrons remaining inside the bubble formed that have not been fully blown out. Operating in this regime has been shown to saturate transverse instabilities \cite{hosingsaturation}. Additionally to the heat-pipe oven, FACET-II can use a high-power laser to ionise the plasma for different gas species such as hydrogen, helium and argon \cite{FACETII}. Hence, we can study the effect of ion motion \cite{ionmotion} on emittance preservation and the trailing-bunch normalised amplitude increase due to the transverse instabilities.

\section{Conclusions}
We have outlined plans for the E302 efficiency--instability experiment at FACET-II. We discuss a technique we will use to quantify the beam's amplitude growth on the imaging spectrometer for the E302 experiment starting in 2024. This method can be used for different configurations of the trailing bunch charge. Therefore, we can analyse the effect of beam loading and efficiency on the beam-breakup instability. As the long-term plan for the E302 experiment, we will also aim to determine the effectiveness of instability mitigation strategies. We will mainly investigate emittance-preserving ion motion and operation in the quasi-linear regime.

\section*{Acknowledgments}
This work was supported by the Research Council of Norway (NFR Grant No. 313770). We acknowledge Sigma2 - the National Infrastructure for High-Performance Computing and
Data Storage in Norway for awarding this project access to the LUMI supercomputer, owned by the EuroHPC Joint Undertaking, hosted by CSC (Finland) and the LUMI consortium.

\section*{References}
\bibliography{main}
\end{document}